\documentclass[sigconf]{acmart}
\AtBeginDocument{%
  \providecommand\BibTeX{{%
    \normalfont B\kern-0.5em{\scshape i\kern-0.25em b}\kern-0.8em\TeX}}}

\copyrightyear{2023} \acmYear{2023} \setcopyright{acmlicensed}\acmConference[GoodIT '23]{ACM International Conference on Information Technology for Social Good}{September 6--8, 2023}{Lisbon, Portugal} \acmBooktitle{ACM International Conference on Information Technology for Social Good (GoodIT '23), September 6--8, 2023, Lisbon, Portugal} \acmPrice{15.00} \acmDOI{10.1145/3582515.3609537} \acmISBN{979-8-4007-0116-0/23/09}

\begin{document}

\title{From Ukraine to the World: Using LinkedIn Data to Monitor Professional Migration from Ukraine}

\author{Margherita Bertè}
\affiliation{
  \institution{ISI Foundation}
  \city{Turin}
  \country{Italy}}
\email{margherita.bert@gmail.it}
\orcid{0000-0003-4372-9734}

\author{Daniela Paolotti}
\affiliation{
  \institution{ISI Foundation}
  \city{Turin}
  \country{Italy}}
\email{daniela.paolotti@gmail.com}
\orcid{0000-0003-1356-3470}

\author{Kyriaki Kalimeri}
\affiliation{
  \institution{ISI Foundation}
  \city{Turin}
  \country{Italy}}
\email{kyriaki.kalimeri@isi.it}
\orcid{0000-0001-8068-5916}

\renewcommand{\shortauthors}{Bertè M. et al.}
\renewcommand{\shorttitle}{Monitoring Ukrainian Professional Migration}

\begin{abstract}
Highly skilled professionals' forced migration from Ukraine was triggered by the conflict in Ukraine in 2014 and amplified by the Russian invasion in 2022. Here, we utilize LinkedIn estimates and official refugee data from the World Bank and the United Nations Refugee Agency, to understand which are the main pull factors that drive the decision-making process of the host country. 
We identify an ongoing and escalating exodus of educated individuals, largely drawn to Poland and Germany, and underscore the crucial role of pre-existing networks in shaping these migration flows. Key findings include a strong correlation between LinkedIn's estimates of highly educated Ukrainian displaced people and official UN refugee statistics, pointing to the significance of prior relationships with Ukraine in determining migration destinations. We train a series of multilinear regression models and the SHAP method revealing that the existence of a support network is the most critical factor in choosing a destination country, while distance is less important.
Our main findings show that the migration patterns of Ukraine's highly skilled workforce, and their impact on both the origin and host countries, are largely influenced by pre-existing networks and communities. This insight can inform strategies to tackle the economic challenges posed by this loss of talent and maximize the benefits of such migration for both Ukraine and the receiving nations.

\end{abstract}

\ccsdesc[500]{Human-centered computing~Collaborative and social computing}
\ccsdesc[300]{Applied computing~Sociology}
\ccsdesc[500]{Information systems~Social advertising}

\keywords{migration, Ukraine crisis, 
LinkedIn Advertising Platform,
social networks, pull factors}


\received{31 May 2023}
\received[revised]{20 July 2023}
\received[accepted]{21 July 2023}

\maketitle

\section{Introduction} 
The global movement of highly skilled professionals has gained significant attention in recent years, as the international labor market continues to witness the exodus of talented individuals from their home countries. Ukraine, a nation rich in intellectual capital, has experienced a notable surge in emigration, particularly among its highly skilled workforce, following the recent conflicts that unfolded in the country~\cite{berte2023monitoring, State2014, Heo2023}.

The phenomenon of migration from Ukraine to other European countries, in particular for employment reasons, is not new~\cite{Mulska2021, Libanova2019}.
The 2014 conflict in Ukraine resulted in a protracted and devastating crisis that significantly impacted various aspects of the nation's social, economic, and political landscape. 
The nature and dynamics of internal migration among Ukrainians have been significantly altered due to the annexation of Crimea in March 2014~\cite{Jaroszewicz2015} affecting around 5 million individuals~\cite{OHCHR2014}. 
By the end of 2021, the number of Ukrainian citizens who had obtained a valid residence permit and were authorized to reside in the EU had reached 1.57 million. This group of people accounted for the third largest group of non-EU citizens residing in the EU~\cite{Eurostat_UA_2023, OECD2022}.

More recently, in 2022 the Russian invasion of Ukraine led to further political instability, economic challenges, and a sense of uncertainty, catalyzing the migration of highly skilled professionals seeking better opportunities and stability abroad.
Referring to the Ukrainian forced migration wave that started on February 2022  Lloyd et al~\cite{Lloyd2022} state that ``the most recent exodus from Ukraine should be seen as the continuation of a legacy of human insecurity in the country".

The phenomenon of skilled migration, often referred to as ``brain drain," has far-reaching implications for both the sending and receiving countries. For Ukraine, the loss of highly skilled professionals poses significant challenges to its economic development, as the country struggles to retain its human capital and nurture innovation within its borders. Conversely, the countries that attract these talented individuals benefit from their expertise, fueling economic growth and fostering innovation in various sectors.
Here, we draw on a combination of quantitative and qualitative research methodologies, utilizing mainly the LinkedIn platform, shown to provide critical demographic estimates that facilitate the study of various topics relating to labor dynamics~\cite{verkroost2020tracking, Kashyap2021}.
We also incorporate Meta's social connectedness index, and sociodemographic data from official sources to comprehensively analyze the migration trends and their implications. 
We explore the role of major determining factors, including economic opportunities, political stability, but also social networks in shaping the decision-making process of highly skilled professionals for the choice of the destination country.
By gaining a deeper understanding of the motivations and consequences of the migration of highly skilled professionals from Ukraine to foreign countries, this research aims to inform policy interventions and initiatives that can help address the challenges faced by Ukraine while harnessing the potential of its human capital. 
We hope that this study will contribute to the development of evidence-based strategies that foster sustainable economic growth, improve social stability and create an environment conducive to the retention and global collaboration of highly skilled professionals.

This paper aims to investigate and predict not the migration flow patterns of highly skilled professionals from Ukraine to European countries, but the determinant factors in the choice of the host country, with a specific focus on the post-war period.
In light of the humanitarian emergency, it is important to estimate not only the proportions of the displacement but also the migration preferences to be able to support host countries in prioritizing aid and efforts. 

\section{Related work}

Scientists endeavored to enhance the official estimates of various countries by incorporating diverse data sources, enabling them to anticipate and gather figures more promptly and flexibly. Many of these approaches had previously demonstrated consistency and significance in different subject areas.

The ``Operational Data Portal" (UNHCR~\cite{UNHCR2023}) provided by the United Nations High Commissioner for Refugees offers reports on ongoing emergencies and presents the latest estimates obtained from authoritative sources. However, the collection and centralization of such precise data pose significant challenges, especially in emergencies. The need for up-to-date data has become indispensable in our rapidly progressing world, often taken for granted. The current challenge lies in ethically and constructively integrating slow yet reliable data from authoritative sources with more volatile, easily accessible data.
Leveraging on traditional macro-data from the European  Statistical Office and national statistics, Mulska et al.~\cite{Mulska2021} highlighted the migration patterns of Ukraine's population after the aggravation of the political situation due to the Russian aggression of 2014. 
They further showed how several European countries introduced active policies to attract foreign labor in particular student-age populations. Such policies established the countries' attractiveness for external migrants, largely determined by medical and demographic stability, better labor market values and employment indicators, living standards, and economic growth.

In addressing this challenge, the inclusion of non-traditional data support and integration becomes crucial. For studying migration and mobility patterns, various data sources are utilized, including geolocated Twitter~\cite{Mazzoli2020}, Facebook advertising data~\cite{Zagheni2017, Mazzoli2020, Fatehkia2022, Palotti2020, Capozzi2021}, ORCID data~\cite{Urbinati2019}, LinkedIn ad estimates~\cite{Vieira2022, Alexander2022, Heo2023}, and bibliometric measures~\cite{Zhao2023}. 
Sirbu et al.~\cite{Sirbu2021} examined the three phases of migration, namely, the journey, stay, and return, by reviewing existing research and enriching insights through the utilization of both big data and traditional data sources.
Focusing on the recent Ukrainian conflict, Juric~\cite{Juric2022}, demonstrated how Google Trends can provide early indications of the besieged Ukrainian population using simple keywords like ``border crossing''. 
Leasure et al.~\cite{Leasure2022} assessed the population displacement by comparing Facebook advertising estimates data collected before and after the Russian invasion on February 24, 2022, while Minora et al.~\cite{Minora2022} examined migratory flows to external countries.
Minora et al.~\cite{Minora2023} employing  Meta's Social Connectedness Index (SCI) and Ukrainian stocks, demonstrated the role of the ``network effect" in displacement trajectories of Ukrainian people during the Russian invasion. 

Here we contribute by a thorough study of the driving factors that influence the choice of the destination country, which despite the efforts, remains an open research question.
Putting together traditional and non-traditional data, i.e. LinkedIn estimates, Meta's SCI, and national migration surveys, we provide a holistic view of the ongoing migration phenomenon, while identifying the major pull factors.
Our aim is, therefore, not to predict flow numbers, but to help reveal trends in the choice of destination country combining both digital and traditional data.

\section{Data Collection}\label{sec:data} 

\paragraph{Official Population and Migration Estimates}
We obtained the official population data for each country from the World Bank report (WB~\cite{WB2023}) for 2021\footnote{The 2021 report is the most recent available collection}.
Additionally, we retrieved data from the UN Refugee Agency~\cite{UNHCR2023}, which tracks the migratory flows throughout the world and supplies aid and support.
It monitors the crisis in Ukraine after the Russian invasion providing an up-to-date share of the official data of Ukrainian refugees provided by the authorities in different European countries. The official monitoring site claims that the reported numbers are estimates and regarding the collection method, states: \textit{when an official estimate is not available, the figure provided corresponds to the sum of registrations for Temporary Protection or similar national protection scheme and the number of asylum applications lodged by refugees from Ukraine}.
For our study, we considered the estimates reported on 13 March 2023 for all European countries under investigation, except the United Kingdom (data of 12/04/2023).
In addition, for the United States and Canada, the estimates were reported on 24/02/2023 by NBC News~\cite{NBC2023} and 17/03/2023 by the Canadian government~\cite{Canada2023}, respectively.
Data are not reported with gender or age resolution, rather, the total estimates are provided per country.

\paragraph{Socioeconomic Indicators} 
Inspired by the work of ~\cite{Mulska2021} and the method of~\cite{Goglia2022}, we considered a series of socioeconomic metrics. 
In particular, we include information about the distance as reported in the GeoDist database~\cite{Mayer2012}, and a country's wealth in terms of the Gross Domestic Product (GDP) expressed in US\$, provided by the World Bank~\cite{WB_gdp_2023}.
Regarding safety, we considered data from the  Institute for Economics \& Peace (IEP\footnote{\url{https://www.economicsandpeace.org/research/}}), which released the Global Peace Index (GPI)~\cite{GPI22}. The GPI covers 163 countries, using 23 qualitative and quantitative indicators, and measures the state of peace across three domains: the level of societal Safety and Security; the extent of Ongoing Domestic and International Conflict; and the degree of Militarisation. The scores for each indicator are normalized on a scale of 1 to 5, where qualitative indicators are ranked into five groups and quantitative indicators are scored from 1 to 5, with the third decimal place. A weight for each area is applied when computing the final score.
To measure the levels of connection and the presence of existing ties between the host country and Ukraine, we used the META Social Connectedness Index (SCI,~\cite{SCI2023}) computed in October 2021. 
Specifically, it measures the relative probability that two individuals across two locations are friends with each other on Facebook.    
    
\paragraph{LinkedIn ad estimates}

We collected the sizes of the LinkedIn audiences that attended a university in Ukraine but that are associated with a geographical location different from Ukraine, following the methodology proposed by Viera et al.~\cite{Vieira2022} for mobility assessment using LinkedIn audiences. 
Querying the ad campaign manager via the official application programming interface (API)\footnote{We were based on the open source code by Lucio Melito \url{https://worldbank.github.io/connectivity\_mapping/intro.html}}, we obtained estimates in line with the sociodemographic criteria of the audience of choice. In case the estimated value is lower than 300 members, the platform indicates 0 as a result.
Our focus is on the movements of the Ukrainian population with a university degree, so firstly we obtained a list of the Ukrainian universities~\cite{Wikipedia2022} and aligned those schools to the ones present in the LinkedIn platform (Member School feature).
Then we queried by location (38 nearby countries), gender (female or male~\footnote{Gender is only provided in a binary form.}), age range (18-24, 25-34, 35-54, 55+, NO AGE),  member school, i.e., the university that a member claims to have attended.
For each combination of gender and age range, we obtain the number of LinkedIn members who attended a University in Ukraine and now work or study in a different country.
Finally, we also queried for each host country targeting only by location, getting the size of the LinkedIn community for each destination.
Our data collection captures all estimates of LinkedIn members dislocated at the time of data collection (27/02/2023), not just the estimates of those who have moved since the displacement caused by Russia's invasion on February 24, 2022.     

\section{Methods}

To compare the collected estimates among the host countries, we must first consider the penetration of LinkedIn per country.
To assess this, we obtained the LinkedIn estimated number of users per country $c$ without age and gender distinction ($\textsf{Total population}$ $\textsf{number LinkedIn}(c)$). For the ground-truth data, ($\textsf{Total population}$ $\textsf{number WB}(c)$), we used the figures reported by the WB in 2021 regarding the population estimates.

We define the LinkedIn penetration measure in each country $c$ as:
\begin{equation}                    \label{eq:LinkedIn_scaling_factor}
    \textsf{LinkedIn penetration}(c) = \frac{\textsf{Total population number LinkedIn}(c)}{\textsf{Total population number WB}(c)}
\end{equation}

Finally, we scaled the LinkedIn estimate collected in each host country $c$ ($\textsf{LinkedIn estimate scaled}(c)$) 
 for all queries by dividing them by this factor. 

\begin{equation}                    \label{eq:LinkedIn_scaled}
    \textsf{LinkedIn estimate scaled}(c) = \frac{\textsf{LinkedIn estimate}(c)}{\textsf{LinkedIn penetration}(c)}
\end{equation}

To understand the evolution of estimates in time, we repeated the entire data collection process in five distinct periods (on February 27, 2023; on April 28, 2023; on May 5, 2023; on May 12, 2023, and on May 19, 2023). 
We then computed the difference of two estimates for each country $c$ between two periods $i$ and $j$, with $j \ge i$ (chronological order). 

\begin{multline}
\label{eq:LinkedIn_diff}
    {\textsf{LinkedIn estimate scaled difference}(c)}_{i,j} =\\{\textsf{LinkedIn estimate scaled}(c)}_j - {\textsf{LinkedIn estimate scaled}(c)}_i
\end{multline}
where ${\textsf{LinkedIn estimate scaled(c)}}_i$ is the collected estimates for the country $c$ at time $i$.
To frame the magnitude of the change and to compare them among countries, we computed the percentage difference with respect to the older estimate.

\begin{multline}                  \label{eq:LinkedIn_percentage_diff}
    {\textsf{LinkedIn estimate scaled percentage difference} (c)}_{i,j} = \\\frac{{\textsf{LinkedIn estimate scaled difference}(c)}_{i,j}}{{\textsf{LinkedIn estimate scaled}(c)}_i}
\end{multline}

\begin{table}[hb!]
    \caption{Summary of the full experimental schema. For each combination of the driving factors (independent variables), we aim to predict the number of refugees estimated by UNHCR.}
    \label{tab:experimental}
    \begin{tabular}{ l l l } 
     \toprule
        Model name &  Predictors \\
     \midrule
Model 1 & LinkedIn estimates\\
Model 2 & SCI \\
Model 3 & distance\\
Model 4 & GPI\\
Model 5 & GDP\\
Model 6 & LinkedIn estimates, SCI\\
Model 7 & LinkedIn estimates, distance\\
Model 8 & LinkedIn estimates, GPI\\
Model 9 & LinkedIn estimates, GDP\\
Model 10 &  LinkedIn estimates, SCI, distance\\
Model 11 &  LinkedIn estimates, SCI, GPI\\
Model 12 & LinkedIn estimates, SCI, GDP\\
Model 13 & LinkedIn estimates, SCI, GDP, GPI\\
Model 14 & LinkedIn estimates, SCI, distance, GPI\\

     \bottomrule
\end{tabular}
\end{table}

\begin{figure*}[ht!]

    \includegraphics[width=0.8\textwidth]{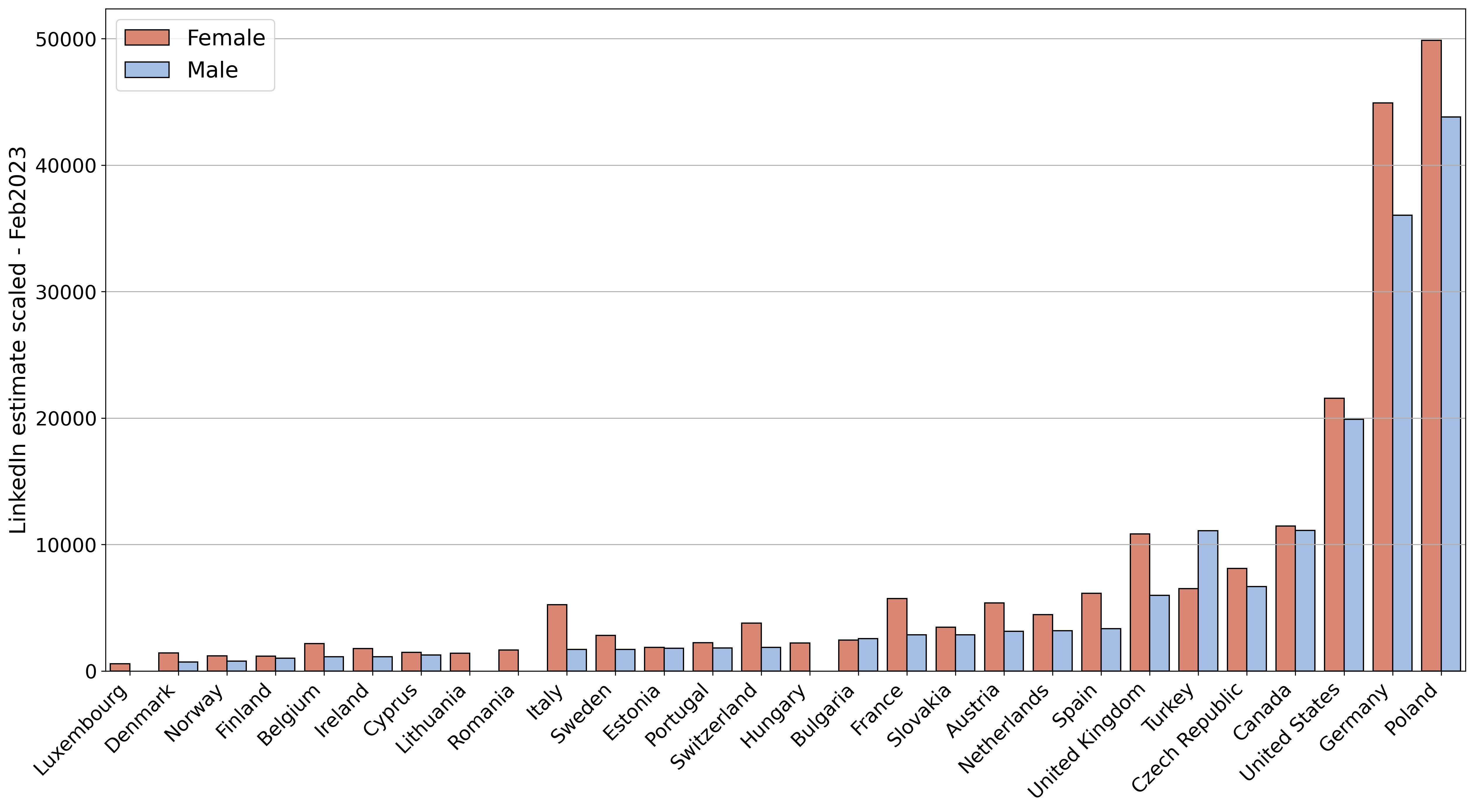}
    \centering
    \caption[LinkedIn estimates rescaled by host country and gender.]{LinkedIn estimates of Ukrainian tertiary educated people by the host country and gender. The estimates are re-scaled by the LinkedIn penetration of each host country.}
    \label{fig:LinkedIn_gender_country}

\end{figure*}

\paragraph{Modelling Migration}
To identify the main pull factors influencing the choice of the destination, we postulated the question as a classification task.
We trained a series of multi-linear regression models (see Table~\ref{tab:experimental}) where for each host country (38 in total) we aimed to predict the number of refugees estimated by UNHCR. To avoid multicollinearity among the independent variables, we used different sets of variables.
The independent variables include the LinkedIn estimates, the geographic distance from the host country to Ukraine, the SCI between Ukraine and the host country in 2021, the GDP of 2021, and the GPI of the host countries in 2022.
Finally, we normalize the features to directly compare their regression coefficients.
Namely, as a common procedure, we computed the standard score of sample $x$ as $x_{norm} = (x - m) / s$, where $m$ is the mean of the samples, and $s$ is the standard deviation of the samples.
To further highlight the importance of certain factors in driving the migrants' host country preference, we apply the SHAP (SHapley Additive exPlanations) method. The goal of SHAP is to explain the result of an instance by computing the contribution of each feature to the prediction to increase the transparency and interpretability of the machine learning model.

\begin{figure*}[th!]
     \includegraphics[width=0.75\textwidth]{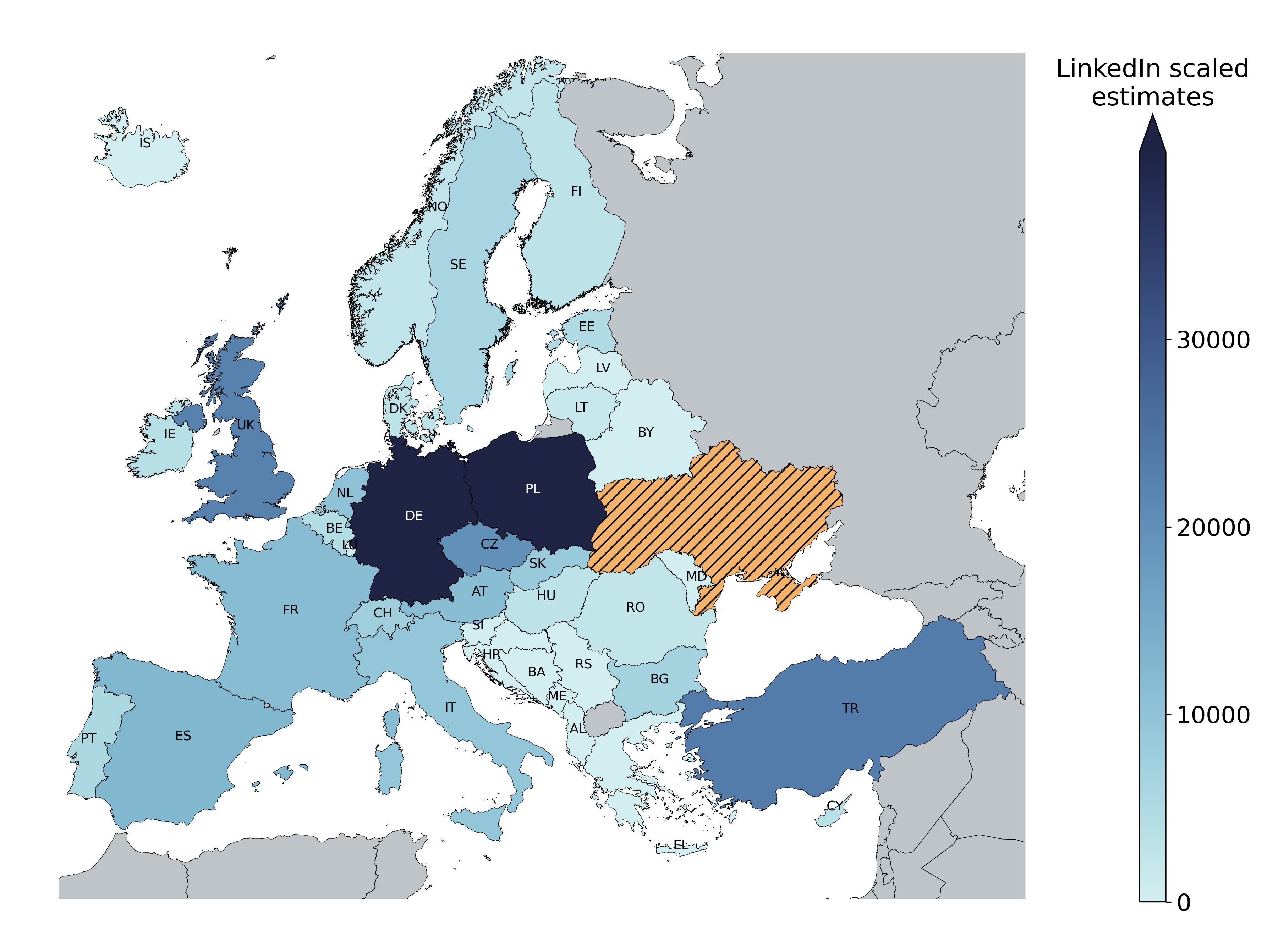}
    \centering
    \caption[Geographic representation of LinkedIn scaled estimates among the European countries.]{Geographic representation with choropleth map of LinkedIn scaled estimates among the European countries. Ukraine is depicted in orange stripes, grey territories do not appear within our analysis.}
    \label{fig:LinkedIn_geopandas}

\end{figure*}

\section{Results \& Discussion}

Overall we collected data from 38 countries; we have positive estimates for females for 28 countries, while for male mobility we have positive estimates for 24 countries.
The two most represented age groups are the ones covering the active labor market, namely, 25-34 and 35-54 years old. 
Taking age into account, overall, we have the highest estimates for women compared to men, having attended a Ukrainian university and now residing in a different country.
Exceptions are the 35-54 age group in the United States, Canada, Poland, the Czech Republic, and Germany, and the 55+ age group in Canada and the United States.
Finally, in Turkey and Bulgaria, more men than women are estimated, regardless of age group.
Among the 10 most preferred destination countries, the top eight out of ten and the tenth are common for men and women (Poland, Germany, the United States, Canada, the United Kingdom, the Czech Republic, Turkey, Spain, and Austria). 
Figure~\ref{fig:LinkedIn_gender_country} depicts the Ukrainian migrants per European country by gender, while Figure~\ref{fig:LinkedIn_geopandas} shows the geographic distribution of migration with darker areas to be highly preferred by the migrants.

Firstly, we assess the representativity of our data,  showing that LinkedIn estimates can be used to support official data in monitoring migration trends.
To do so, we compared LinkedIn data (without gender distinction) by the host country with the official estimates reported by UNHCR about the number of Ukrainian refugees across Europe.

\begin{figure*}[ht!]
    \includegraphics[width=.75\textwidth]{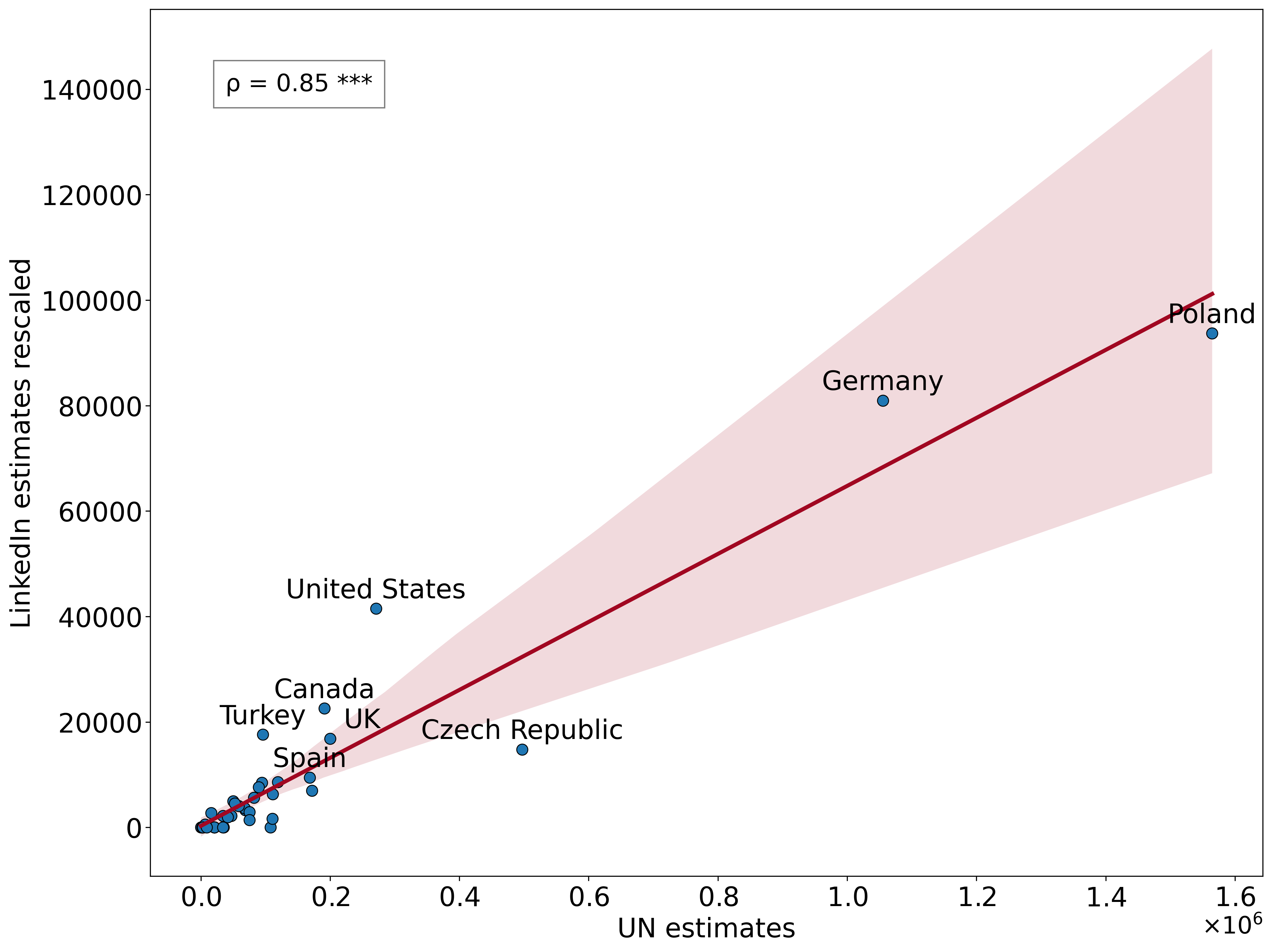}
    \centering
    \caption[LinkedIn estimates rescaled by host country compared with UN estimates.]{LinkedIn estimates of Ukrainian tertiary educated people by host country versus UN official estimates of Ukrainian migrants by host countries. The Spearman correlation coefficient $\rho$ and its significance are reported. Notation for the significance is the following: *** $\text{p-value} \leq 0.001$, ** $\text{p-value} \leq 0.01$, * $\text{p-value} \leq 0.05$. The dark red line shows a regression line with 95\% confidence.}
    \label{fig:LinkedIn_validation_UN_linkedin}
\end{figure*}

We find a significant positive Spearman\footnote{We opted for the Spearman correlation coefficient, due to the presence of outliers with much higher values than in other countries (Germany and Poland).} correlation coefficient $\rho$ ($\rho = 0.85$, $\text{p-value} \leq 0.001$) between LinkedIn estimates of tertiary educated Ukrainian migrants and UNHCR official figures of Ukrainian refugees by the host country (see Figure~\ref{fig:LinkedIn_validation_UN_linkedin}), implying that LinkedIn's estimates of highly educated Ukrainians abroad are a good indicator of the actual number of Ukrainian refugees in a country. This can be justified by the results of the ongoing survey led by UNHCR for Regional Protection Profiling \& Monitoring (from October 2022 onwards), which points out that most refugees have a high level of education (78\%) and pre-war employment (60\%), categories well covered by LinkedIn.
Additionally, Ukraine has a much higher percentage of highly educated people than the European average (in the 30-34 age group 65\% for women, 51\% for men in 2021~\cite{EUROSTAT2023}) and LinkedIn is a fairly widespread social media with penetration of 8.3\% of the population~\cite{Datareportal2022}. 
Hence, the LinkedIn ad estimates act as a valid proxy for the displacement of highly skilled professionals but are also a good proxy for the general population. 

Facilitating the attainment of gainful employment that aligns with the educational and professional qualifications of refugees not only enhances their self-sufficiency but also contributes to the economic growth of the host community.
To further investigate the outcome of the migration, we decided to collect data using LinkedIn's Seniority feature as a target in the data collection to understand in the target countries which roles are most frequently filled by the highly educated Ukrainian migrants.
We calculated for each country, grouped by gender, the percentage of workers in the different roles.
Looking at the different distributions for each possible seniority level (Unpaid, Training, Entry, Senior, Manager, Director, VP, CxO, Partner, Owner), it emerges that the median for women is 51\% for Entry level, 42\% Senior, for all other roles the median distribution is zero. Similarly, for men, 48\% Entry level, 40\% Senior and 3\% hold the role of Director.
This outcome could potentially be attributed to data sparsity, meaning the occurrences wherein the count of individuals affiliated with a specific role and that received education from a Ukrainian university falls below the minimum threshold of 300, resulting in a value of 0. With sparse availability of data it is difficult to discern the actual distribution of roles held. 
We observe a notable presence of Senior and Entry level positions for both genders, with a nearly equal distribution among the audience. However, since information regarding users' prior occupations is unavailable, it remains uncertain whether their current roles align with their previous professional experiences.

\subsection{Temporal Evolution of Migration Patterns} 
The two time windows examined to assess the stability of the estimates are ten weeks and a week long, respectively.
Inspired by the study of Rama et al.~\cite{Rama2020} who thoroughly assessed the temporal stability of Facebook estimates especially in under-represented populations, we assessed the stability of our estimates for the entire period of analysis. 
We considered two periods: the first considers the entire data collection window spanning over 10 weeks from 27/02/2023 to 12/05/2023. The second one examines the fluctuations over one week within the entire data collection period.
For the weekly analysis, we breakdown the consecutive one-week periods:
(i) from 28/04/2023 to 05/05/2023, (ii) from 05/05/2023 to 12/05/2023 and (iii) 12/05/2023 to 19/05/2023. 
The one-weeks periods aim at assessing the stability of the approach, while the ten-week one indicates the overall trend.

\begin{figure*}[th!]
    \includegraphics[width=0.85\textwidth]{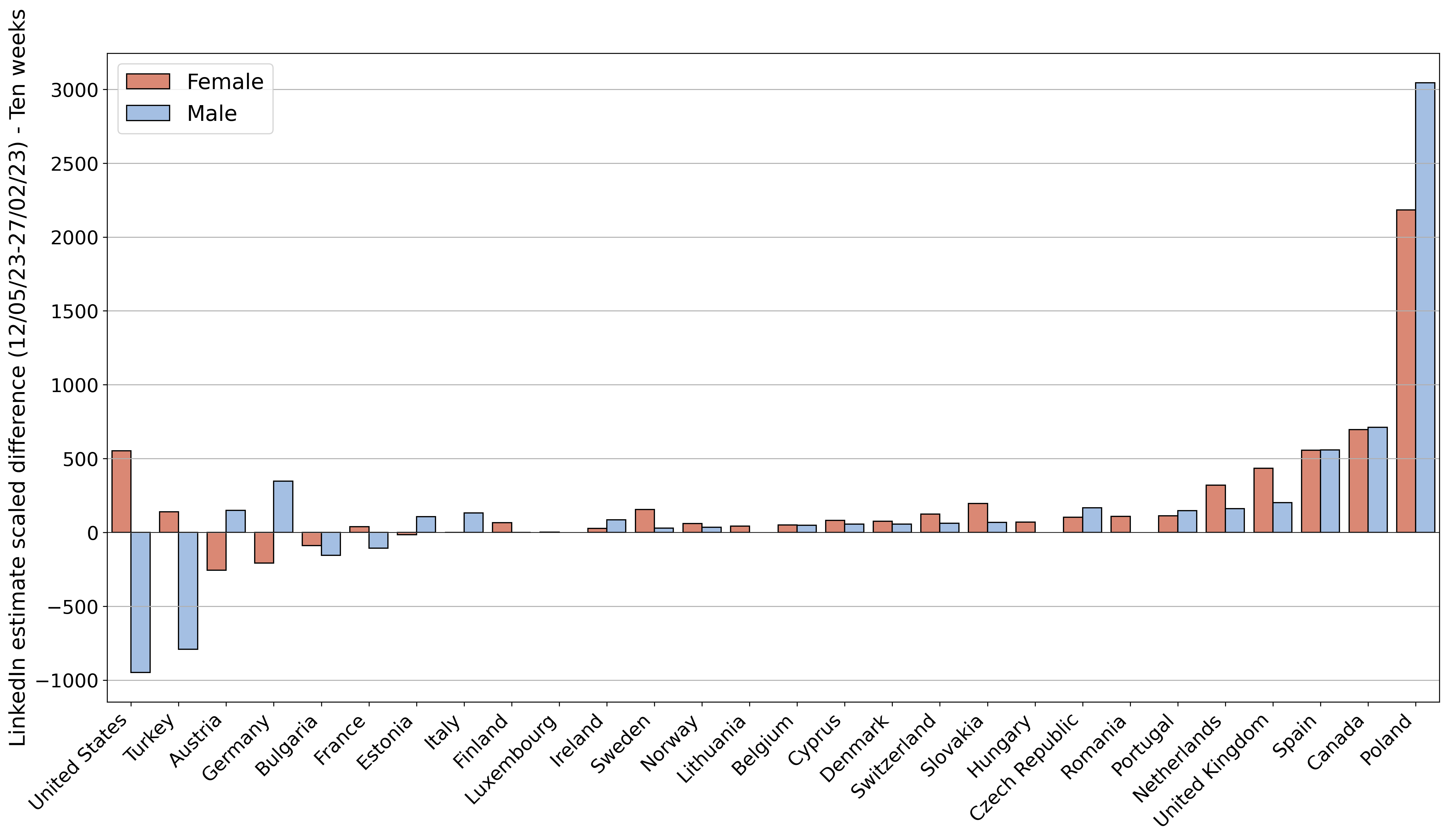}
    \centering
    \caption[Ten weeks difference of LinkedIn estimates scaled by host country and gender.]{Differences of LinkedIn re-scaled estimates of Ukrainian tertiary educated people by the host country and gender between the data collected on May 12, 2023 and February 27, 2023.}    \label{fig:LinkedIn_tenweek_diff}
\end{figure*}

Over the entire period of ten weeks, the percentage changes become more significant with about 75\% of the countries under 6.5\% and over 1\% for both genders.
The estimates show a clear and stable increasing migration trend in time; Figure~\ref{fig:LinkedIn_tenweek_diff} depicts the trend of the ${\textsf{LinkedIn estimate scaled difference}(c)}_{i,j}$ per country $c$ during the entire period (Eq.~\ref{eq:LinkedIn_diff} with $i=27/02/2023$ and $j=12/05/2023$). 

\begin{figure*}[ht!]
    \includegraphics[width=0.85\textwidth]{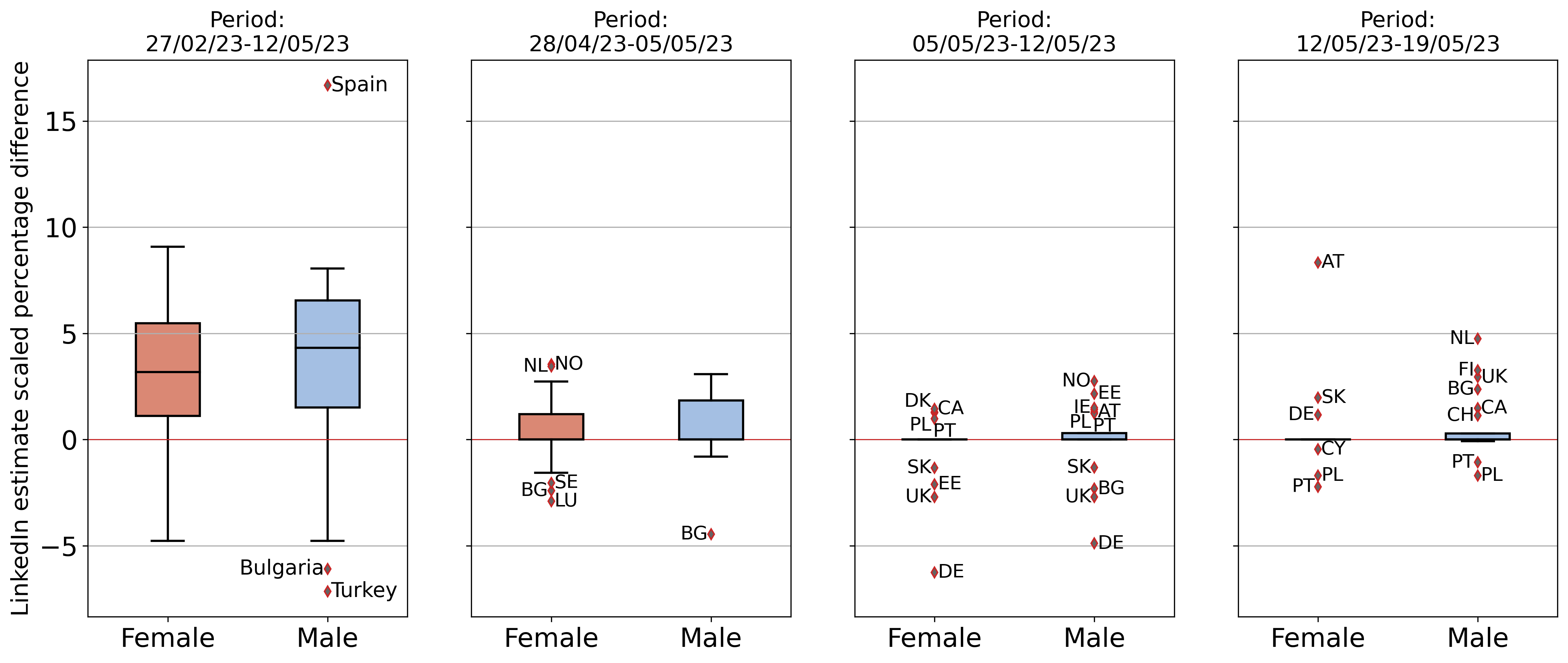}
    \centering
    \caption[Boxplots of percentage differences in LinkedIn estimates rescaled by the host country and gender.]{Boxplots of the percentage differences by gender and temporal window of the LinkedIn re-scaled estimates of Ukrainian tertiary educated people. We highlight the outlier countries for each case. The first boxplot refers to the ten-weeks data collection period (from 27/02/23 to 12/05/23) while the consecutive boxplots refer to the one-week resolution data, showing the stability of the estimates (date 28/04/23-05/05/23, 05/05/23-12/05/23, 12/05/23-19/05/23)}
    \label{fig:LinkedIn_diff_box_perc}
\end{figure*}

In Figure~\ref{fig:LinkedIn_diff_box_perc} we compare the distributions of ${\textsf{LinkedIn}}$ ${\textsf{estimate}}$ ${\textsf{scaled percentage difference (c)}_{i,j}}$ for the two different time windows by gender across the destination countries $c$.
The last three plots of Figure~\ref{fig:LinkedIn_diff_box_perc} depicts the distributions of the percentage differences for three one-week periods compared with the first boxplot related to the ten-week one. 
We notice that for both the time-windows the estimates increased with a higher peak for men in Spain in the ten weeks.

The second and third weeks have smaller inlier values with respect to the first week, on the other hand, the last ones have bigger outliers, for example, in the second week Germany has a change of -6.25\% for women and -4.89\% for men.
This can be explained because the magnitude of \textsf{Total population number LinkedIn (Germany)} changed from 150.000 on May 05, 2023 to 160.000 on May 12, 2023. This growth increased the \textsf{LinkedIn penetration (Germany)} factor in the second week, and hence, the scaled estimates become significantly lower.

\subsection{Modelling Migration} 

\begin{table*}[ht!]
    \caption{Summary of results for seven regression models predicting Ukrainian migrants in the different destination countries. We report the model fit and regression coefficients for normalized measures using different sets of independent variables. Notation for the significance is the following: *** $\text{p-value} \leq 0.001$, ** $\text{p-value} \leq 0.01$, * $\text{p-value} \leq 0.05$}
    \label{tab:model_coef_multimodel}
    \centering
    \begin{tabular}{ l l l l l l  l l} 
     \toprule
     & \multicolumn{5}{c}{Coefficients} & \multicolumn{2}{c}{Model Fit} \\
     \cmidrule(lr){2-6}\cmidrule(lr){7-8}
         & LinkedIn estimates  & SCI & Distance & GPI & GDP & ${R}^2_{\text{adjusted}}$ & F-statistic \\
     \midrule
        Model 1 & 0.95***  &          &          &          &          & 0.91     & 329.2***    \\
        Model 2 &          & 0.32*    &          &          &          & 0.08     & 4.1*        \\
        Model 6 & 0.92***  &  0.18*** &          &          &          & 0.93     & 242.5***    \\
        Model 7 & 0.99***  &          &  -0.2**  &          &          & 0.94     & 265.3***    \\
        Model 8 & 0.96***  &          &          & -0.13**  &          & 0.91     & 194.7***    \\
        Model 9 & 1.04***  &          &          &          & -0.22*** & 0.94     & 296.9***    \\
        Model 10 & 0.96***  & 0.12**   &-0.15**   &          &          & 0.95     & 216.0***    \\
        Model 11 & 0.93***  &  0.20*** &          &  -0.15** &          & 0.95     & 243.3***    \\
        Model 12 & 1***     &  0.13*** &          &          & -0.19*** & 0.96     & 277.5***    \\
        Model 13 & 0.99***  & 0.16***  &          & -0.10**  & -0.14*** & 0.97     & 248.8***    \\
        Model 14 & 0.96***  & 0.15***  & -0.13*** & -0.14*** &          & 0.97     & 265.1***    \\
     \bottomrule
\end{tabular}
\end{table*}

To address our main research question, namely, which are the main pull factors driving the migrants' choice of the host country, we trained a series of predictive models. 
We consider a set of factors inspired by the existing literature~\cite{Beine2011, Crawley2019, Uebelmesser2013, Manafi2022}, such as the host country's distance from Ukraine, safety (peacefulness), the presence of pre-existing relationships as a support network in the host country, and the wealth of the host country. 
We followed an incremental experimental design. By combining several pull factors, we built multi-linear regression models to predict for each host country the number of Ukrainian refugees estimated by UNHCR.

\begin{figure*}[h!] 
    \centering
    \begin{minipage}{0.47\textwidth}
        \centering
        \includegraphics[width=1\textwidth]{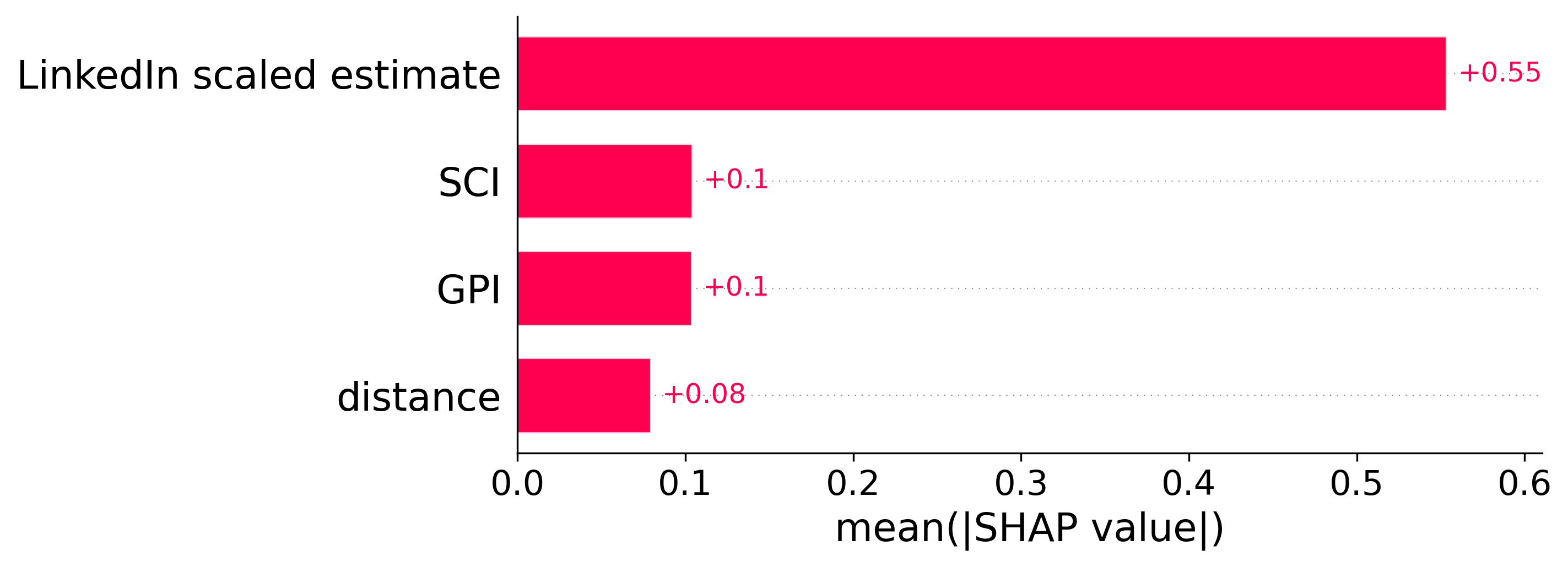}
        \caption[SHAP values]{We report the mean of the absolute SHAP value for each country for Model 14, which has the best fit. The features are ordered from the highest to the lowest effect on the prediction.}
        \label{fig:shap_mean}
    \end{minipage}\hfill
    \begin{minipage}{0.47\textwidth}
        \centering
        \includegraphics[width=1\textwidth]{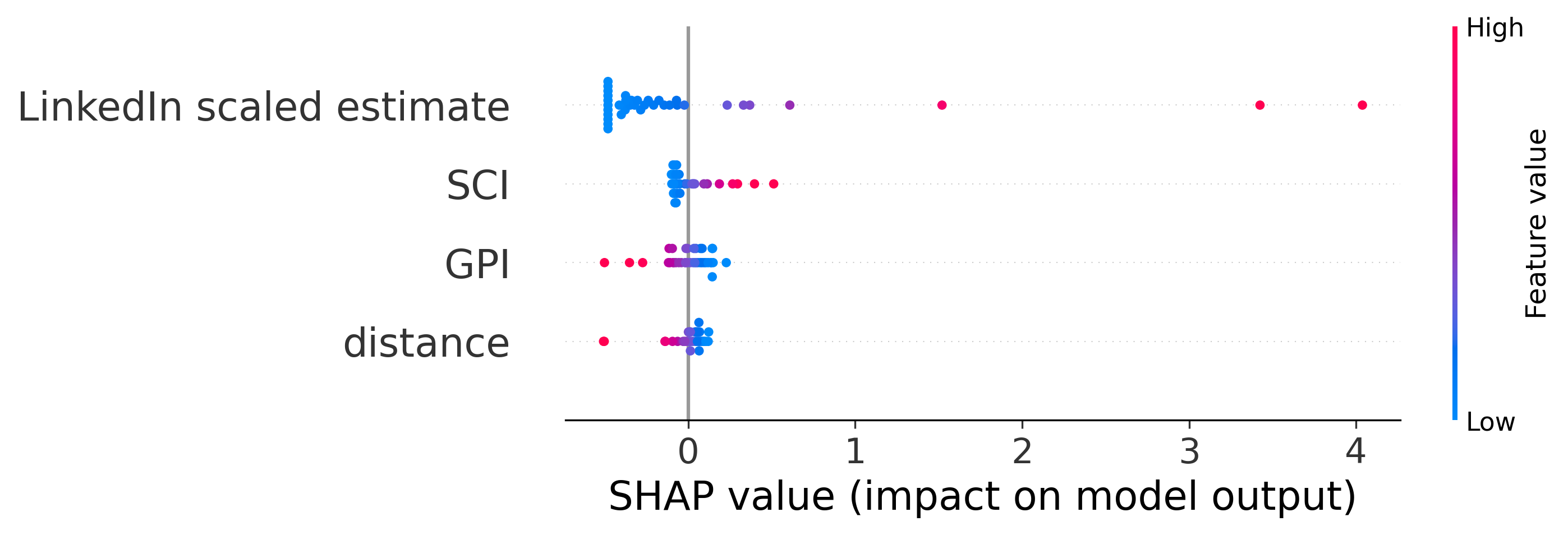}
        \caption[SHAP values local]{SHAP values for all the countries in Model 14. The features are ordered from the highest to the lowest effect on the prediction. High feature values are in magenta and low in blue.}
        \label{fig:shap_single}
    \end{minipage}        
\end{figure*}

Table~\ref{tab:model_coef_multimodel} reports the coefficients obtained for each set of normalized variables with significant coefficients for each trained model.
We observe that the most important feature is the LinkedIn scaled estimates, which confirm the trends of the migration flows reported by UNHCR.
Migration, being a complex phenomenon, requires the examination of a wide range of pull factors to better understand the decision-making process around the choice of the host country. From the predictive models built, we observe that the LinkedIn estimates alone are highly explanatory of the migration flows (Model 1 in Table~\ref{tab:model_coef_multimodel}), and Model 2 has a significant coefficient for the SCI even alone, indicating that people mainly choose the host country based on their social network.
The model that best describes the variation of the migration flows is the one with the largest number of variables (Model 14 in Table~\ref{tab:model_coef_multimodel}).
In this model, the SCI is the second most important feature considering the magnitude of the coefficients.
Regarding distance and GPI, the negative signs of the coefficients are coherent with the meaning of the measure: the countries are less attractive when the two measures are higher, meaning that the closer a country is the less safe is considered.
Finally, in Models 9, 12 and 13 the coefficient for GDP is negative, indicating that economic prosperity is less considered with respect to factors like safety or social network relationships of the people.
GDP incidentally correlates with distance from Ukraine (the more distant countries are also richer) with countries like US and Canada which boost this behavior.

To fully assess the importance of each feature, we computed SHAP values for the best performing model (Model 14).
We present the average of the absolute SHAP values for all countries (Figure~\ref{fig:shap_mean}), confirming that the two most important features are the LinkedIn estimates and the SCI.
In Figure~\ref{fig:shap_single} are displayed the SHAP values for each country.
We observe that the variables LinkedIn estimates and the SCI have a high contribution when their values are positive, and a low contribution on negative values.
The situation is reversed for GPI and distance, which have a higher contribution when their values are negative.

\subsection{Limitations}

In light of the aforementioned potentials of our approach, is crucial to recognize and address the inherent limitations associated with the utilization of non-traditional data sources~\cite{urbinati2020young}, as LinkedIn advertising estimates.
Especially when dealing with vulnerable populations digital data may reveal sensitive demographic patterns  \cite{beiro2022fairness}.
The measures derived from LinkedIn are likely to overestimate the number of tertiary-educated migrants due to the method's limitations. Indeed, the information taken into account regarding universities is declared by the user and cannot be verified. Further users may be counted more than once if they attended more than one university in Ukraine.
Nevertheless, inferring residence based on the university attended provides valuable insight since previous location data are unavailable from the LinkedIn ad platform. 
It is also important to notice that the collected estimates provide a snapshot of the situation at the time of collection. We access data only after February 2022 precluding comparison of earlier migration flows. 
Conversely, official UN figures quantify the total number of refugees since the Russian invasion in February 2022 but lack demographic information, hindering demographic comparisons with LinkedIn data. 
Finally, the UN reported more than 2,85 million Ukrainian refugees in Russia on October 2022~\cite{UNHCR2023}. 
However, Ukrainian refugees in Russia are not investigated in this work since LinkedIn's audience location setting cannot be configured to Russia due to policy restrictions imposed by the Office of Foreign Assets Control (OFAC) of the US Department of the Treasury administers which publishes a list of prohibited trade-sanctioned countries~\cite{OFAC2023}, as reported by LinkedIn in~\cite{LinkedIn2023}.

\section{Conclusion}

The deterioration of relations between Ukraine and Russia escalated into conflict in 2014 and intensified in 2022 with Russia's invasion of Ukraine. As a result, the Ukrainian population suffered significant migration flows in search of stability and better job opportunities, leading to one of the largest displaced workforces in Europe. The Russian invasion in February 2022 has made this need more urgent, strengthening migration flows. 
It is, therefore, imperative that host countries can direct them correctly and have a centralized view of the situation. The UNHCR is collecting estimates provided by the different authorities, struggling to capture the current situation due to the very nature of the problem.
For this reason, non-traditional data sources have been used to support the official estimates.
Our work fits into this context, using estimates from the LinkedIn ad platform to detect the number of people who have moved from Ukraine to another country. To collect the data, we limited our sample to members with tertiary education, using the status of the universities attended by the LinkedIn member as a target. Our collection estimates correlate strongly with official UN estimates and reveal trends in destination choices, with Poland and Germany as the preferred destinations. To investigate potential pull factors in country choice, we developed several linear regression models starting from LinkedIn estimates and indices for factors that by literature are spot as pull factors. The most important features are LinkedIn estimates and the 2021 Facebook Social Connectedness Index, which measures ties between individuals living in different countries. These findings confirm the importance of the presence of a community in the destination country in shaping the destination decision-making process.
In contrast, distance plays a less significant role in migration patterns, which aligns with the nature of the collected estimates primarily representing settled migrants. The urgency to escape conflict swiftly is no longer the primary driver, shifting the focus toward other factors influencing destination choices. However, the observed phenomenon remains ongoing, as evidenced by the moderate yet consistent increase in variations over the longer period, in contrast to the minor fluctuations during the shorter period, indicating stability in the estimates.
To conclude, although our data is restricted to LinkedIn members with tertiary education, we believe that for a highly educated country such as Ukraine, the communities sizes of highly educated people spread in other countries can be a good indicator for tracking flows and may be part of the pull factor that attracts new refugees.


\begin{acks}
    The authors gratefully acknowledge the support from the Lagrange Project of the Institute for Scientific Interchange Foundation (ISI Foundation) funded by Fondazione Cassa di Risparmio di Torino (Fondazione CRT).
\end{acks}


\bibliographystyle{ACM-Reference-Format}
\bibliography{From_ukraine_to_the_world}


\end{document}